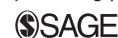

# Enhanced Fluorescence in a Scattering Medium

Nathan A. Giauque, Callum E. Flowerday, and Steven R. Goates

## Abstract

Often only small amounts of sample are available for spectroscopic analytical determinations. This work investigates the enhancement of signal in columns packed with silica particles. We propose that silica particles cause the light to scatter through the column, effectively increasing optical path length. Packed columns are shown to be effective with fluorescence spectroscopy, but results were inconclusive with absorbance spectroscopy.

## Introduction

Sample analysis often requires not only low detection limits but being able to deal with small sample volumes. A variety of approaches have been taken to achieve this for a range of applications.[1–4] As an example, smaller sample volumes allow reduced dead volumes in flowing systems and thereby limit the spread of bands flowing through a detection cell.[5]

We observed enhanced detection in a previous series of studies[6–8] of the flow of supercritical fluids in microcapillary columns (250 μm i.d.) packed with 5 μm diameter silica particles. These studies involved measurements of fluid density using Raman spectroscopy along the length of a column. To our surprise, the Raman signal from microcapillary columns packed with the silica spheres was greater than that from microcapillaries that had no packing material, despite the reduced sample volume in the packed columns. In these studies, we also used laser-induced fluorescence to interrogate the elution of analytes through the column and were able to detect the analytes on the column even with the much-reduced volume. We concluded that signal enhancement occurred because of the multiple reflections of the excitation beam that occurred in the packed column, effectively increasing its optical path length.[6] This effect is sometimes called photon diffusion. Applications of photon diffusion have been used in the study of a variety of materials.[9–12]

In this study, we have tried to more quantitatively examine signal enhancement in cells packed with silica spheres for fluorescence and absorption measurements.

## Experimental

For the fluorescence experiments, the fluorescent dye rhodamine B base (Sigma-Aldrich) dissolved in high-performance liquid chromatography-grade methanol (Sigma-Aldrich) was employed. The earlier chromatography experiments, discussed above, were performed on microcapillaries packed with a supercritical fluid slurry packing method.[13] However, because preparing such packed capillaries requires specialized equipment and some skill, in this current study, we explored a simpler, more accessible approach using melting point tubes (Kimax, 0.9–1.1 mm × 90 mm) and capillary tubes (Pyrex, 0.8–1.1 mm × 100 mm). The melting point tubes were packed with Mallinckrodt 60–200 mesh (75–200 μm), porous silica gel (hereafter referred to as unmodified silica), or with 10 μm porous spherical silica particles with an octadecyl-

Department of Chemistry and Biochemistry, Brigham Young University, Provo, USA

**Corresponding author:**
Steven R. Goates, Department of Chemistry and Biochemistry, Brigham Young University, Provo, UT 84602, USA.
Email: sgoates@byu.edu

(C₁₈) bonded phase from Phenomenex. Capillary tubes were packed with $10\,\mu m$ $C_{18}$-modified porous silica particles from Millipore Sigma. Tubes were packed by filling them through a micropipette tip used as a funnel, tapping the particles down, and centrifuging the tubes at 6000 rpm for half an hour. Solution was added to the melting point tubes with a syringe and centrifuged down into the tube. For the experiments with the capillary tubes, solution was continuously pumped through with a peristaltic pump, as described below. To keep the packing from being pushed out, the end of the capillary tube was melted partially closed and cotton wool was added at the bottom.

The overall experimental setup is shown in Fig. 1. Fluorescence was excited with 1 W of 514.5 nm light from an a Lexel argon ion laser (Lexel). The insets in Figs. 1a and 1b show the illumination of open and packed cells by the laser beam. No focusing lens was used with the excitation beam; the beam diameter at the cell was about 2.5 mm, so the beam overfilled the cell. An aluminum block was milled to allow the tube to reproducibly slide in vertically. This arrangement kept the positioning of the cells fixed with respect to the laser beam and a monochromator, located about 12.7 cm away. No collection lens was used between the cell and the monochromator. A mask was placed at the front of the cell holder to block fluorescence excited by laser light scattered outside of the irradiated area. Greater signal could have been achieved by moving the cell holder closer to the monochromator, but our goal was not to achieve the greatest signal; rather, it was to compare signal of packed and unpacked cells under reproducibly similar conditions, where illumination of the slit was not affected by small changes in cell position or imaging of the cell.

Most experiments were done using a 0.33 m monochromator (Instruments SA, HR320) and a photomultiplier tube (PMT) (bialkali, Thorn EMI, operated at 1100 V). Experiments with melting point tubes were later repeated to check reproducibility; in these experiments, a 0.303 m monochromator (Andor Technology, Shamrock SR-3031-A) and a charge-coupled device (Andor Technology, iDus 420) were used. Both monochromators had an *f*-number of about five. In all experiments, a 532 nm longpass edge filter (Semrock) was placed between the testing block and the monochromator to reduce scattered laser light.

Initial tests were done with the melting point tubes, closed on one end, filled with the dye solution. Measurements were made on unpacked tubes, tubes packed with unmodified silica, and tubes packed with the $C_{18}$-bonded phase particles. Tubes were loaded in one shot to avoid any inadvertent increase in total amount of dye in the observation volume due to the dye being retained at the silica surface. The bonded phase should have further limited adherence of the dye to the particles. However, to reduce the possibility that observed enhancement was due to concentration of the dye on the packing as the cells were filled, instead of to the photon diffusion effect, a set of experiments were performed with a flowing system with capillary tubes, which allowed the signal to be observed to rise and fall with the introduction of sample into the flowing stream. The flowing setup also facilitated running through a range of concentrations repeatedly. A peristaltic pump (Perkin Elmer) with a stated stability of better than 1.5% was used to flow solution through the capillary tubes. Initially, a flow rate of 3 mL/min was used, but at that rate, the tubing with the packed tubes leaked, so the flow rate was reduced to 1.5 mL/min. No dependence of signal on flow rate was observed. Dilute solutions were made in steps within a range of $2.22 \times 10^{-8}$ to $6.94 \times 10^{-10}$ M. Each solution was tested with and without silica packing, and between runs, methanol was used to flush the column. Measurements were made starting with the lowest concentration of rhodamine B and working up to the higher concentrations. After each introduction of the sample, the signal rose smoothly to a constant level for a minute or more until the inlet tube was removed from the sample. After removing the inlet tube from the sample and placing it again in methanol, the signal fell smoothly back to a constant baseline.

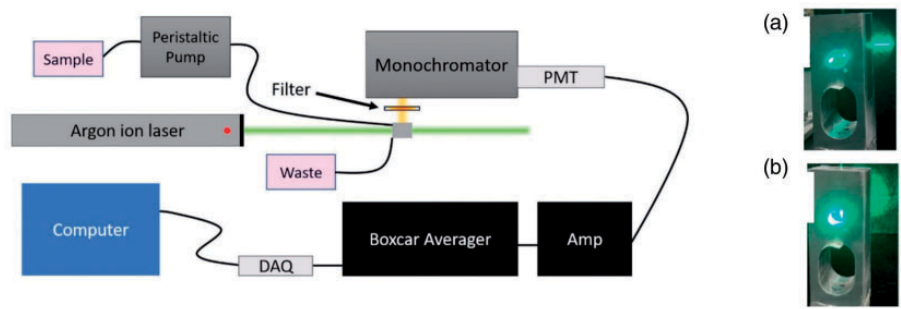

**Figure 1.** Experimental setup. A peristaltic pump was used in for the flowing experiments with the capillary tubes; it was not present in the non-flowing experiments with the melting point tubes. Insets are photographic images showing illumination of the melting point tubes with (a) no packing and (b) packing with unmodified silica.

Tests of possible enhancement of absorption were performed with a frequency-doubled neodymium-doped yttrium aluminum garnet (Nd:YAG) laser (Laser Quantum, Opus) at 532 nm and an aqueous solution of cobalt nitrate. Transmitted light was collected with a ball lens and focused with another lens onto the monochromator entrance slit.

## Results and Discussion

As shown in Fig. 2, the signal from sealed melting point tubes packed with unmodified silica shows a distinct increase in signal over that from tubes with no packing, even without being adjusted for the smaller volume in the packed tubes.

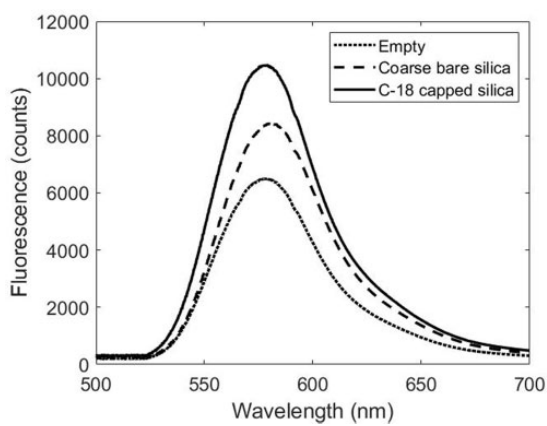

**Figure 2.** Comparison of fluorescence of rhodamine B base in melting point tubes that were unpacked (empty), packed with unmodified silica gel (coarse bare), and packed with 10 μm modified silica ($C_{18}$ capped silica).

The increase in signal was even higher with the $C_{18}$-modified silica. We hypothesize that this greater increase with the modified silica is because of the smaller particle size and increased packing obtained with the modified silica. Repetition of these measurements with new tubes and packing gave reproducible results. The signal from the tubes with unmodified silica shows a shift in the peak maximum of fluorescence; no such shift is observed in the tubes packed with modified silica. This difference in shift is consistent with a strong interaction of the dye with bare silica.

A better test is from the flowing experiments, where a range of concentrations were employed. Results from these measurements are shown in Fig. 3. Over the range of concentrations tested, the signal from the packed tubes was several times that from the empty tubes, again without any adjustment for reduced sample volume. Using the packed columns, we were able to detect higher signal over the unpacked tubes down to the nanomolar level.

The same process of multiple scattering that we ascribe to increased signal in the observation area also causes exciting light to be scattered farther along the cell, as is evident in a comparison of the photographic images in parts (a) and (b) of Fig. 1. Visual inspection of fluorescence from the cell through the cut-off filter also showed that fluorescence was excited outside the directly irradiated area but was much greater in the irradiated area. The mask and the deliberate mismatch of *f*-number between the monochromator and signal discriminate against signal outside the irradiated area. Removing the mask would increase overall signal due to the increased amount of analyte irradiated. This would be at a cost of total sample volume but would still require a smaller volume of analyte sample than in most detection arrangements. The position of the mask was fixed for all experiments, but it could not be precisely placed to eliminate all signals from outside the irradiated area.

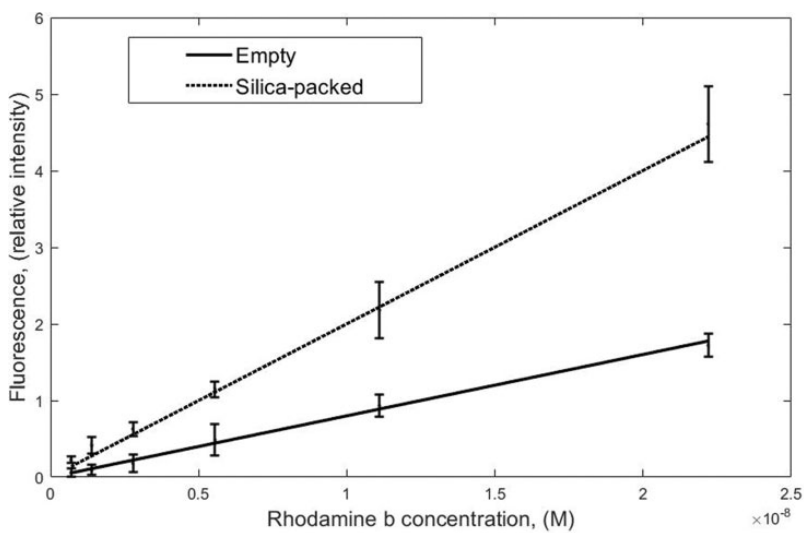

**Figure 3.** Fluorescent response for rhodamine B from an open capillary tube and a tube packed with 10 μm modified silica particles. Error bars represent 95% confidence intervals for four runs; intensity for each run was the average of 200 points after the signal leveled off.

As a related side note, in the earlier studies[6–8] in which we first noticed an effect of packing on signal intensity, measurements were made through windows in the polyimide coating on the microcapillary columns, but the overall optical arrangement was different than in this study, and the signal light was imaged on the slit of a monochromator, so the observation volume was well defined. However, at that time, we did not attempt to quantitatively compare the difference between signals from packed and unpacked columns.

The response from the packed tubes is nonlinear. Nonlinear curves are well known to occur with increasing concentration and path length for fluorescence, but it is curious that the line here curves up. This behavior was reproducibly observed for four runs in which new tubes were used in each run, and fresh test solutions were prepared for each pair of runs. We do not have an explanation for the upward curvature, but we note that the effect of an absorbing species on photon diffusion can be complex.[9]

Results for absorption were not reproducible. In some cases, enhancement in absorbance was observed, but in many cases, the absorbance was lower. We think the problem is that too little light overall was transmitted through the packed columns with our setup. It is possible that signal enhancement can be achieved but only with much better collection of light passing through the column, such as with an integrating sphere. As it is, however, we do not see packed columns as a practical approach for most absorption applications, especially because much work has already been done on improving absorption detection, e.g., Aiello and McLaren.[5]

## Conclusion

A silica-packed column can increase the fluorescence signal, allowing for analysis of smaller volumes than without packing. This makes feasible on-column detection in micro- or capillary-column chromatography or allows a smaller size of post-column detection cell that would reduce extra-column band broadening. In general, when only a small amount of sample is available, this scheme allows measurement of the sample without the drawback of having to dilute it to fill a cell, or it allows detection of a sample that must be diluted for other reasons.

We have demonstrated such small volume detection for fluorescence and previously for Raman signals,[6] but not for absorption. It could be worthwhile to investigate the use of other packing materials.

## Acknowledgments

We thank Jared Dickson for assistance with fluorescence measurements.

## Declaration of Conflicting Interests

The author(s) declared no potential conflicts of interest with respect to the research, authorship, and/or publication of this article.

## Funding

Student salaries and supplies were provided by the College of Physical and Mathematical Sciences at Brigham Young University.

## ORCID iD

Steven R. Goates  https://orcid.org/0000-0002-4520-095X